\documentclass[journal]{IEEEtran}
\usepackage{cite}

\ifCLASSINFOpdf
   \usepackage[pdftex]{graphicx}
\else
\fi
\usepackage{amsmath}

\usepackage{xcolor}

\begin{document}
%
\title{Experimental validation of ultra-shortened 3D ﬁnite element models for frequency-domain analyses of three-core armored cables}
%
%
%

\author{Juan Carlos del-Pino-López, and 
	Pedro Cruz-Romero,~\IEEEmembership{Member,~IEEE}
\thanks{Manuscript received September xx, 20xx; revised December xx, 20xx.}
\thanks{This research was funded by FEDER/Ministerio de Ciencia e Innovación – Agencia Estatal de Investigación under the project ENE2017-89669-R and by the Universidad de Sevilla (VI PPIT-US) under grant 2018/00000740.}
\thanks{The authors are with the Electrical Engineering Department, Universidad de Sevilla, Camino de los Descubrimientos s/n, 41092 Sevilla, Spain (e-mail: vaisat@us.es, plcruz@us.es).}
}

%
%

\markboth{Accepted for publication in IEEE Transactions on Power Delivery (author's version). Citation: DOI 10.1109/TPWRD.2022.3158870}%
{Shell \MakeLowercase{\textit{et al.}}: Bare Demo of IEEEtran.cls for IEEE Journals}
%

\IEEEpubid{© 2022 IEEE. Personal use is permitted, but republication/redistribution requires IEEE permission. See https://www.ieee.org/publications/rights/index.html for more information.}


\maketitle

\begin{abstract}
Recently, large offshore wind power plants have been installed far from the shore, using long HVAC three-core armored cables to export power. Its high capacitance may contribute to the appearance of unwanted phenomena, such as overvoltages or resonances at low frequencies. To adequately assess these problems, detailed and reliable cable models are required to develop time-domain/frequency-domain analyses on this type of cables. This paper presents, for the first time in the literature, an assessment on the performance of 3D finite element method-based (3D-FEM) models for developing frequency-domain analyses on three-core armored cables, confronting simulation results with experimental measurements found in the literature for three real cables. To this aim, a simplified ultra-shortened 3D-FEM model is proposed to reduce the simulation time during frequency sweeps, through which relevant aspects never analyzed before with frequency-domain 3D-FEM simulations are addressed, such as total losses, induced sheath current, magnetic field around the power cable,  positive and zero sequence harmonic impedances, as well as resonant frequencies. Also, a time-domain example derived from the frequency-domain analysis is provided. Remarkable results are obtained when comparing computed values and measurements, presenting the simplified ultra-shortened 3D-FEM model as a valuable tool for the frequency-domain analysis of these cables.
\end{abstract}


%
\IEEEpeerreviewmaketitle

\section{Introduction}
%
%
%
%
\IEEEPARstart{D}{uring} the last decades, the number of offshore wind power plants (OWPPs) has significantly increased, and this trend is expected to continue in the coming years \cite{cigre147}. This is forcing the technology to new limits, with larger OWPPs placed far from the shore \cite{wind} that require longer export cables. This has important consequences for the system, especially when installing long HVAC three-core armored cables (TCACs), since its high capacitance may contribute to the appearance of unwanted phenomena, such as switching or temporary overvoltages \cite{Jensen2014,Jansen2015,Palone2016,TB556}, as well as resonances at low-order harmonic frequencies \cite{TB556,Palone2014,Freijedo2015,Wang2020,Bakhshizadeh2016}. 

To adequately assess these problems on TCACs, detailed and reliable cable models are required to develop time-domain/frequency-domain analyses  \cite{TB556,dasilva2013,Bakhshizadeh2016,Kocewiak2018,Norouzi2019}, where important aspects, like frequency-dependent skin/proximity effects and armor losses, have to be properly considered to avoid costly over-design \cite{Bakhshizadeh2016,Kocewiak2018}. In this sense, during the last decade, the use of numerical simulations, such as the finite element method (FEM), has become one of the most powerful and versatile tools for analyzing TCACs. Thus, 2D-FEM \cite{Bremnes2010b,dasilva2016,Hafner2015a,Furlan2019,Ruixiang2019} models have been extensively employed for obtaining their electrical parameters. However, the complex geometry of this type of cables requires the use of 3D geometries, so that the influence of the armor and three-core twisting is adequately considered. To this aim, important advances have been developed in the last years \cite{Benato2017,Chatzipetros2018a,Willen2019,Sturm2020}, although huge computational resources are required. To overcome this problem, \cite{Pino2018a} proposed a shortened 3D-FEM model by including rotated periodicity as an additional boundary condition. This model was further simplified (without loss of accuracy) in \cite{Pino2020c,Pino2021a,Pino2021b,Cagigal}, reducing the simulation time below 1 minute. This ultra-shortened 3D-FEM model (USM) has been exhaustively faced to experimental measurements for evaluating its performance regarding relevant aspects of TCACs at power frequency, such as the series resistance, inductive reactance, induced sheath current, and zero sequence impedance \cite{Pino2021h}. However, the extent to which the USM can be useful for the analysis of TCACs at harmonic frequencies remains to be determined. In this sense, since these cables are specifically built on a project-by-project basis, it is difficult for the academia to have access to a piece of cable for testing purposes. So, few research studies are available providing experimental measurements derived from a frequency-domain analysis on real TCACs \cite{Arentsen2017,Palone2019,Benato2021}, being simulations tools, like the USM, a valuable alternative for the analysis and design of TCACs, supplementing costly experimental setups.  
 \IEEEpubidadjcol
Having all this in mind, this paper goes a step further by proposing a simplified ultra-shortened 3D-FEM model (SUSM) to reduce the simulation time during frequency-domain analyses. Its performance in the harmonic frequencies is evaluated by replicating previously published experimental studies on a set of three real TCACs of different voltages, cross sections and conductor materials. The relative differences between measurements and simulated values regarding relevant aspects never analyzed before through frequency-domain 3D-FEM simulations are assessed, such as total losses, induced sheath currents, magnetic field (MF) distribution around TCACs,  positive and zero sequence harmonic impedances, as well as resonant frequencies. A very good matching between simulation results and measurements are observed, showing how useful the SUSM can be for the frequency-domain analysis of TCACs. Finally, a time-domain example derived from the frequency-domain analysis is also provided.

\section{SUSM for frequency sweeps}

\begin{figure}[!htb]
	\centering
	\includegraphics[width=5 cm]{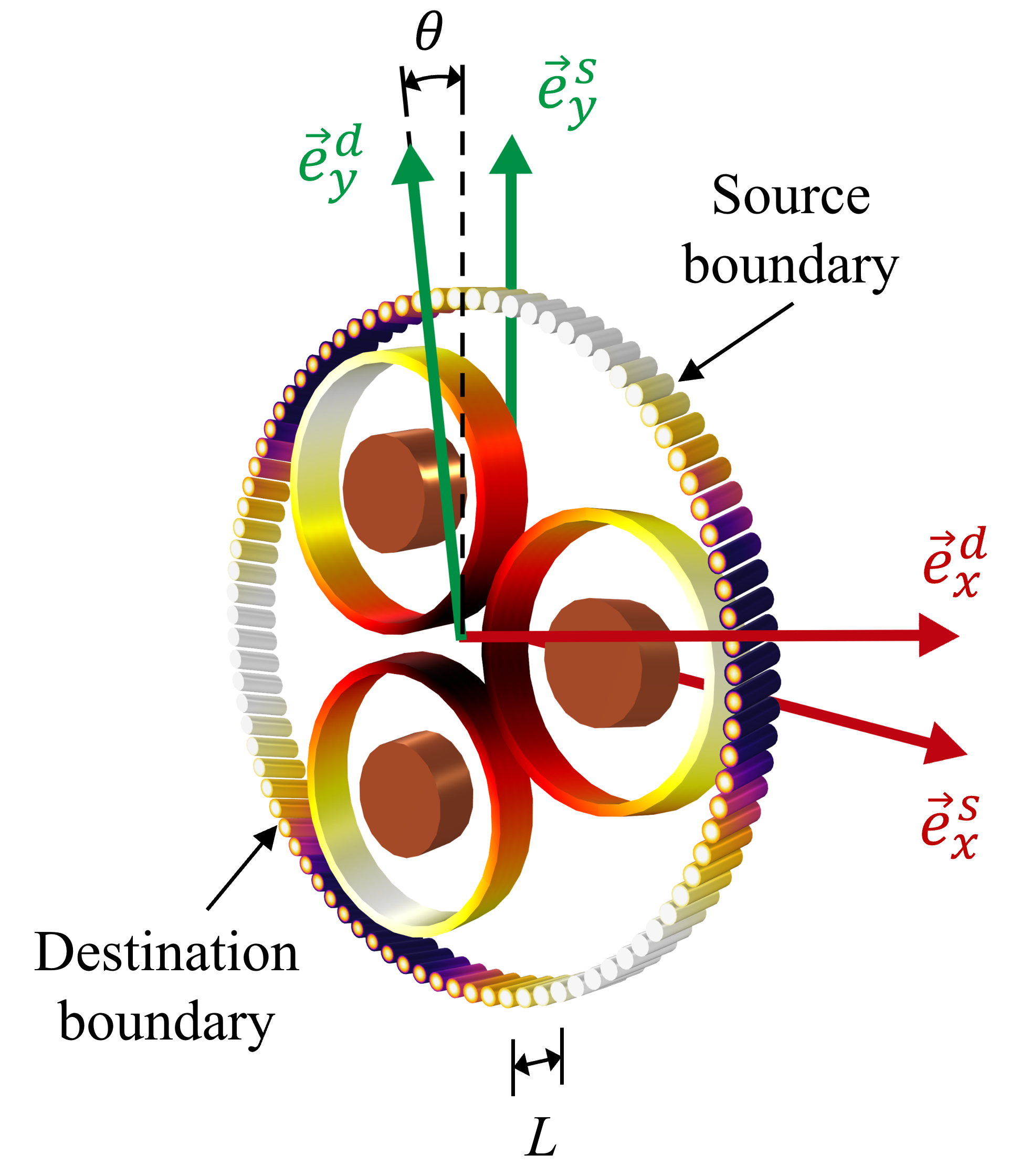}
	\caption{USM: Rotated periodicity and current density distribution in TCACs.\label{fig1}}
\end{figure}

The USM presented in \cite{Pino2020c,Pino2021a,Pino2021b} is based on the symmetries found in the geometry and the electromagnetic field distribution in TCACs, so that rotated periodicity can be applied in just a small slice of the cable (Figure~\ref{fig1}) with a length ($L$) equal to

\begin{equation}
	L=\frac{1}{N\cdot\left|\cfrac{1}{L_c}-\cfrac{1}{L_a}\right|},
\end{equation}\label{eq1_jc}

\noindent where $N$ is the number of armor wires, and $L_a$ and $L_c$ are, respectively, the lay length of armor wires and phase conductors (being $L_a<0$ when they are twisted in different directions (contralay) and $L_a>0$ if twisted in the same direction (unilay)). The relative rotation angle ($\theta$) for properly mapping the~source boundary into the~destination~boundary  (Figure~\ref{fig1}) is 

\begin{equation}
	\theta=2\pi\cfrac{L}{L_c}.
\end{equation}\label{eq2_jc}

\begin{figure}[!htb]
	\centering
	\includegraphics[width=4.5 cm]{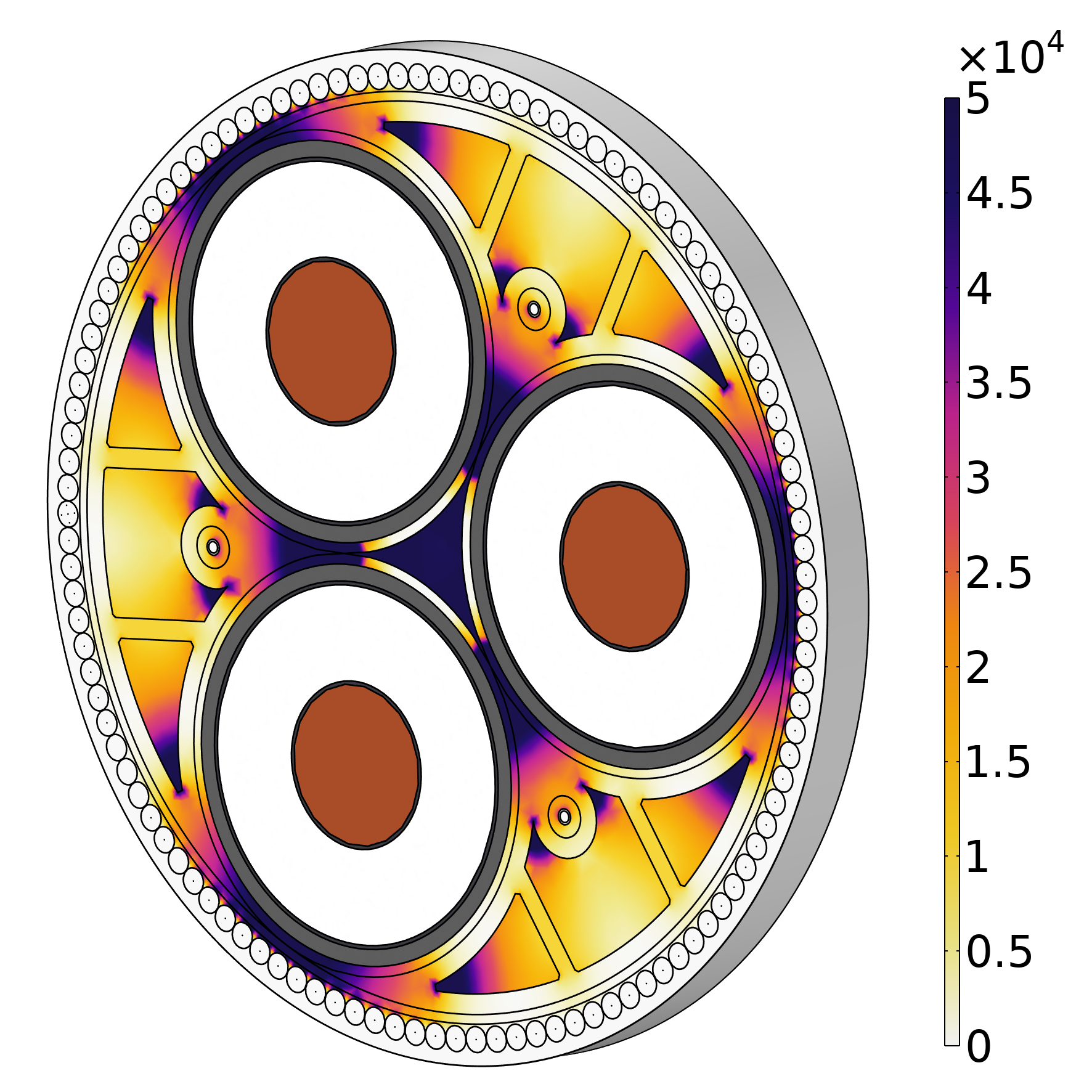}
	\caption{USM: Electric field distribution (V/m) within TCAC's fillers.\label{fig2}}
\end{figure}

This USM (implemented in COMSOL Multiphysics by means of the AC/DC module \cite{comsol}) reduces the simulation time to less than 1 minute \cite{Pino2021a} for a single computation of the MF problem in TCACs, and less than 30 seconds if solving for the electric field (EF). As a consequence, finer meshes can be employed without greatly increasing the computational requirements, so that more detailed 3D geometries can be analyzed (i.e. including semiconductive layers, optic fiber cables or fillers for simulating the EF within TCACs (Figure \ref{fig2}). Moreover, for any frequency, meshes can now be better adjusted for a proper evaluation of the skin depth ($\delta$) \cite{TB531} within all metallic parts in TCACs, especially in the armor wires (Figure \ref{fig1}). Finally, and in contrast to analogous studies \cite{Benato2021}, the USM can employ nonlinear properties for the armor wires.

\begin{figure}[!htb]
	\centering
	\includegraphics[width=8 cm]{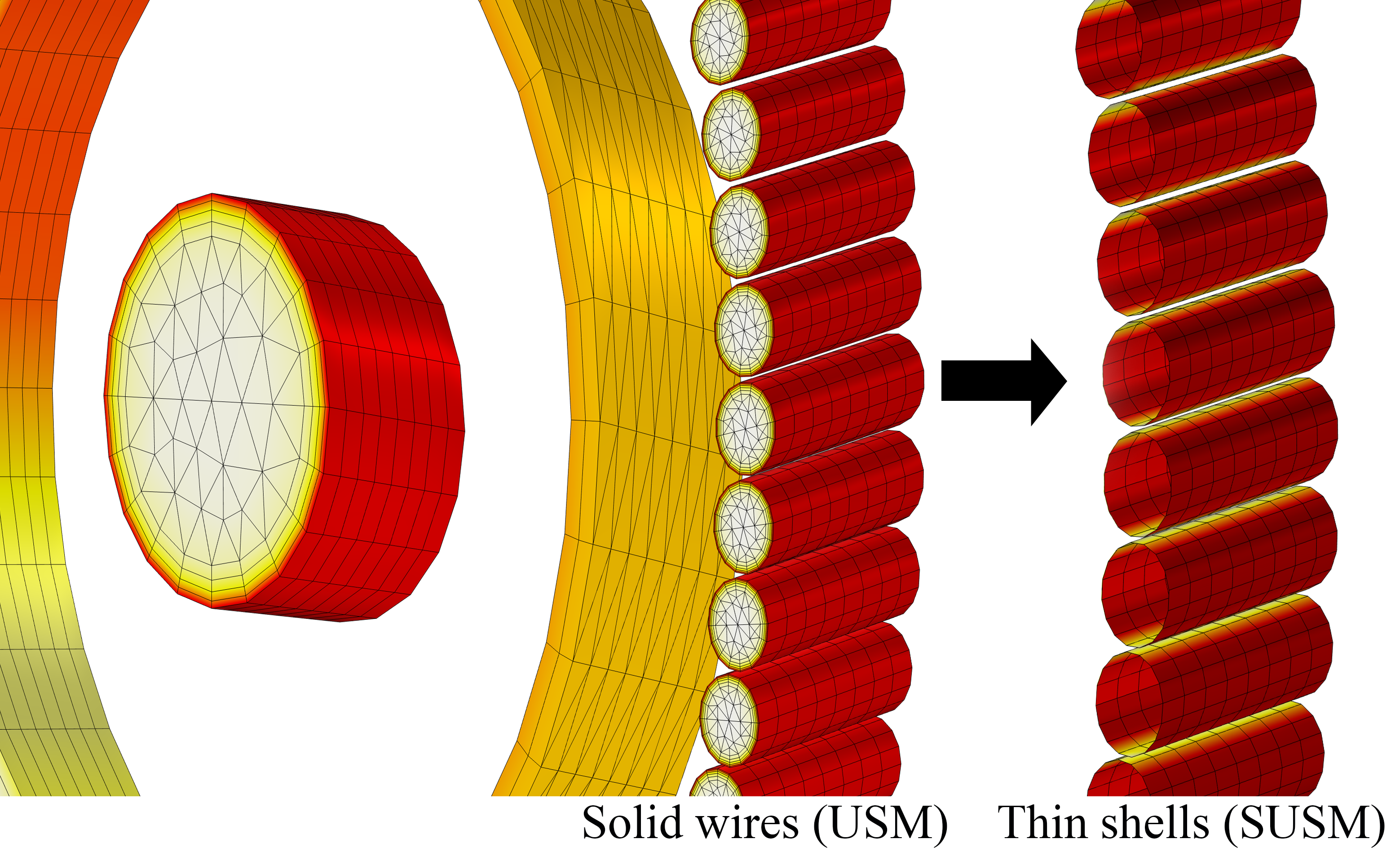}
	\caption{Mesh employed for the armor wires when using the USM or the impedance boundary condition in the SUSM.\label{fig3}}
\end{figure}

Nonetheless, although simulation time is usually short for a single computation, for a frequency sweep analysis the number of simulations to be performed may lead to a total simulation time longer than desired. Therefore, it is of interest to have a further reduction of computation time, especially for cable design optimization purposes. In this sense, due to the small armor wire diameter ($d_a$) and its material properties, it is usual that $\delta < d_a/2$ for frequencies higher than 30 Hz. Therefore, for harmonic frequencies $\delta$ becomes very small (Figure \ref{fig3}), so that the armor wires can be assumed as thin steel shells where the impedance boundary condition can be applied when solving the MF problem \cite{comsol}. This results in the SUSM proposed in this work, where only a surface mesh is employed in the armor wires  (Figure \ref{fig3}), which represents a significant reduction in the number of unknowns of the MF problem.

Additionally, special caution must be devoted to the size of the simulation domain, especially when computing the zero sequence impedance in actual submarine installations. In this case, the same current is injected through the three conductors, {assuming the current return through the sheaths, armor and the surrounding medium. In the SUSM, the latter is considered as homogeneous and having the properties of the sea} when simulating a submarine cable on its actual location. Following \cite{TB531}, its radius must be, at least, equal to the Carson’s depth, defined for submarine cables as

\begin{equation}
	d_{sea} \cong 503 \sqrt{\frac{\rho_{sea}}{f}},
\end{equation}\label{eq2a}

\noindent where $\rho_{sea}$ is the electrical resistivity of the sea (200 m$\Omega\cdot$m \cite{Benato2021,TB531}). This leads to different radii for the surrounding medium depending on the frequency, which can be far beyond 100 m for $f<50$ Hz. Thus, to limit the size of the 3D-FEM model, $d_{sea}$ is customarily scaled in a shorter distance by applying a non-linear coordinate transformation to the surrounding medium layer, having the effect of stretching it to almost infinity ("infinite element domain" feature of \cite{comsol}). This size is automatically adjusted for each frequency in order to reduce the extent of the model as the frequency increases, and hence reducing the computation time. Also, both the mesh of the surrounding medium and the metallic elements (depending on $\delta$) are optimized for each frequency.

To evaluate the extent to which the SUSM is applicable for the MF analysis of TCACs, some results are summarized in Table \ref{tab1} regarding the positive and zero sequence resistances ($R_+, R_0$) and inductive reactances ($X_+,X_0$), induced sheath current ($I_s$), losses in conductors ($P_c$), sheaths ($P_s$) and armor ($P_a$), and the MF at 0.5 m from a 630 mm$^2$, 220 kV TCAC (cable C2 in Table \ref{tab2}). In particular, they show the relative differences ($\varepsilon$) obtained for different frequencies ($f$) when comparing the results derived from the SUSM with respect to those of the previous USM for frequencies up to 1 MHz.

\begin{table}[!htb]
	\renewcommand{\arraystretch}{1.1}
	\begin{center}
		\caption{Relative differences in the zero and positive sequence electrical parameters, losses and MF provided by the USM and SUSM for the C2 cable of Table \ref{tab2}.}
		\label{tab1}
		\scalebox{0.93}{
			\begin{tabular}{c|*{9}{c}}
				& \multicolumn{7}{c}{Relative differences $\varepsilon$ (\%)}	\\
				$f$ (Hz) &$\varepsilon_{R_0}$ 	&$\varepsilon_{X_0}$ 	&$\varepsilon_{R_+}$ 	&$\varepsilon_{X_+}$ &$\varepsilon_{I_s}$ 	&$\varepsilon_{P_c}$  &$\varepsilon_{P_s}$ 	&$\varepsilon_{P_a}$ &$\varepsilon_{MF}$ \\ \hline 
				0.5		&11	  	&36		&4.5	&5.6	&12		&0.2	&29		&$10^3$	&91	\\
				1		&15		&33		&4.0	&5.2	&11		&0.2	&27		&$10^3$	&73\\
				5		&34		&24		&3.6	&3.6	&8.4	&0.1	&18		&825	&47\\
				10		&32		&22		&3.2	&2.4	&6.0	&0.1	&13		&450	&32\\
				30		&9.5	&11 	&3.2	&2.4	&6.0	&0.1	&13		&450	&8.8\\				
				50		&1.9	&8.3	&2.9	&0.4	&0.2	&0.1	&0.6	&66		&2.8\\
				100		&0.3	&3.0	&0.9	&0.3	&0.2	&0.0	&0.3	&23		&1.9\\
				500		&0.2	&0.1	&0.1	&0.0	&0.1	&0.0	&0.2	&10		&1.2\\
				1000	&0.1	&0.0	&0.0	&0.0	&0.1	&0.0	&0.1	&7.0	&0.5\\
				2000	&0.1	&0.0	&0.0	&0.0	&0.0	&0.0	&0.0	&3.4	&0.4\\
				5000	&0.0	&0.0	&0.0	&0.0	&0.0	&0.0	&0.0	&2.6	&0.3\\
				$10^4$	&0.0	&0.0	&0.0	&0.0	&0.0	&0.0	&0.0	&2.8	&0.3\\
				$5\cdot10^4$	&0.0	&0.0	&0.0	&0.0	&0.0	&0.0	&0.0	&3.6	&0.2\\
				$10^5$	&0.0	&0.0	&0.0	&0.0	&0.0	&0.0	&0.0	&4.6	&0.2\\
				$5\cdot10^5$	&0.0	&0.0	&0.0	&0.0	&0.0	&0.0	&0.0	&4.2	&0.2\\				
				$10^6$	&0.0	&0.0	&0.0	&0.0	&0.0	&0.0	&0.0	&4.5	&0.2\\
																				
			\end{tabular}
		}
	\end{center}
\end{table}

As can be observed, for $R_+$, $X_+$ and $P_c$ the relative differences are very low for all $f$, as well as for $I_s$ and $P_s$ when $f>30$ Hz. However, as expected, greater errors are observed for $P_a$, where acceptable values are only obtained for $f>500$ Hz, although this is mainly due to the small values of $P_a$ for lower frequencies (usually below 2 W/m). Regarding the MF at 0.5 m from the TCAC, differences below 10 \% are obtained for $f>30$ Hz. Eventually, if the TCAC is supplied with a zero sequence current, differences below 10 \% are obtained in $R_0$ when $f>30$ Hz, while similar results are obtained in $X_0$ when $f>50$ Hz. Nonetheless, although the SUSM loses accuracy in a short frequency range ($f<$ 30 Hz), this has a limited impact on the results obtained in typical frequency-domain analyses, as shown in the following sections.
	
On the other hand, the SUSM reduces the number of degrees of freedom in 25 \% (from 730724 to 549404). This is of importance since an extensive frequency sweep noticeably increases the memory requirements and the total simulation time. Thus, the SUSM reduces the total simulation time in about 33 \% compared to the USM (from 388 s to 260 s), so that a frequency sweep can be performed in a computer with less than 16 GB of RAM memory.

\begin{figure}[!htb]
	\centering
	\includegraphics[width=9 cm]{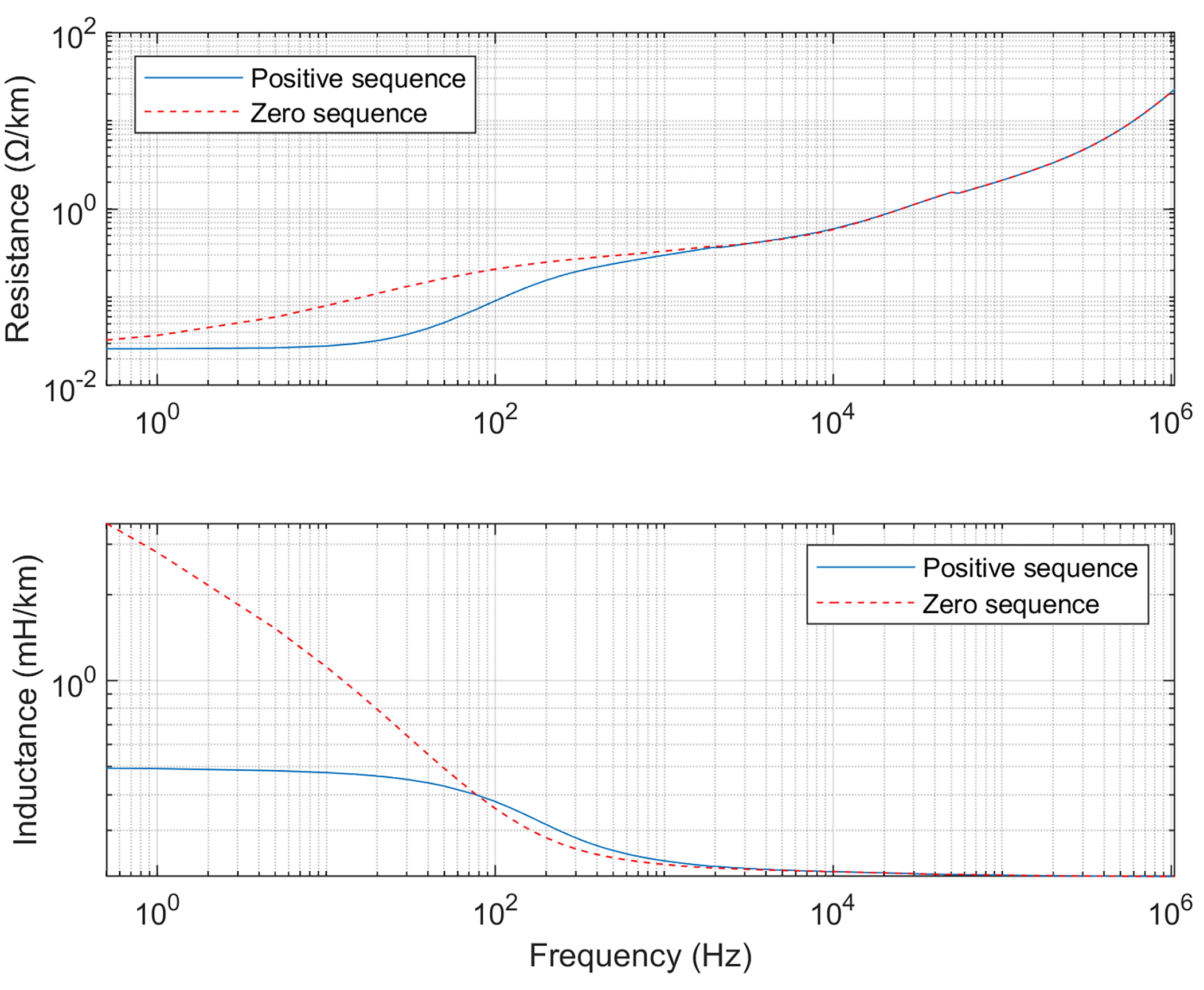}
	\caption{Cable C2 (SB): positive and zero sequence series resistance and inductance as a function of frequency (derived by the SUSM).\label{fig4}}
\end{figure}

Another example of its application can be found in Figure \ref{fig4}, which shows the positive and zero sequence series resistance and inductance when a complete frequency sweep with SUSM is performed in cable C2 (sheaths and armor in solid bonding configuration (SB)), where results are in line with similar studies in TCACs \cite{Patel2016}.

Consequently, for typical frequency-domain analyses the SUSM provides a good balance between accuracy and computational burden in a wide frequency range.

\section{Case studies}
Few studies can be found in the literature providing a complete set of data and experimental measurements obtained during frequency analysis of TCACs. This work is based on the most detailed studies that can be found in the specialized literature for three real TCACs. Table \ref{tab2} summarizes their main dimensions and properties, being $I_{max}$ the rated current, $d_c$ the conductor diameter, $D_s$, $D_{core}$ and $D_a$ the outer diameter of sheaths, power cores and armor, respectively, $e_s$ the sheath thickness, and $d_a$ the armor wire diameter. It should be noted that, for the C1 cable, just a few geometrical (e.g. the lay lengths) or material parameters (the armor permeability) were missing in the references, so they had to be estimated by comparison with similar TCACs or derived from other data provided by the authors. Hence, higher comparison errors are expected for this cable.

\begin{table}[!htb]
		\renewcommand{\arraystretch}{1.1}
		\begin{center}
		\caption{Main dimensions and properties of the three TCACs analyzed.}
		\label{tab2}
			\begin{tabular}{cccc}
				&  \textbf{C1}     & \textbf{C2}   &\textbf{C3}  \\ \hline
				Reference		&\cite{Arentsen2017}	&  \cite{Palone2019,Benato2021} 	& \cite{Hatlo2014b} \\
				$V_r$ (kV)		& 30 		& 220 	   & 132     \\
				$I_{max}$ (A)	 & 200	& 675 	   & 732    \\
				$S_n$ (mm$^2$) 	 & 120	& 630	   & 800\\
				Material	     & Al 	& Cu	   & Cu\\
				$d_c$ (mm)		& $13.4$	& $30.5$   & $35$    \\
				$D_s$ (mm)		& $37$	& $92.1$  	& $87.6$  \\
				$e_s$ (mm)	 	& $1.7$	& $3$      & $3.7$  \\
				$D_{core}$ (mm)	& $41.57$	& $97.3$   & $92.4$  \\
				$D_a$ (mm)		& $99$	& $238.6$ 	& $214.6$ \\
				$d_a$ (mm)	 	& $4$		& $5.6$    & $5.6$  \\
				$N$ 			& $69$	& $120$   	& $114$   \\
				$L_a$ (m)	 	& $1.2$	& $3$     & $3.5$  \\
				$L_c$ (m)	 	& $1$  	& $2.7$  	& $2.8$  \\
				Twist	 		& cont.   	& cont.	   & cont. \\ 
				$T_{amb}$ ($^\circ$C)	& $10$  	& $5$   	& $5$   \\
				$\sigma_c$ (MS/m)	& $39.06$  	 & $53.1$   	& $51$  \\
				$\sigma_s$ (MS/m)	& $3.25$  	& $4.97$   	& $4.50$ \\
				$\sigma_a$ (MS/m)	& $6.9$  	& $5.16$   	& $5.19$\\
				$\mu_r$				& Figure \ref{fig5}a   	& Figure \ref{fig5}a    	& Figure \ref{fig5}a     \\														
				$\varepsilon_r$		& -   	& 2.5    	& -     \\
				tan$\ \delta$		& -   	& Figure \ref{fig5}b    	& -     \\
			\color{red}	$L_{cable}$ (m)		& -   	& \color{red}99650    	& \color{red}80     \\				
				$L$ (mm)		& 8   	& 12    	&14     \\								
			\end{tabular}
	\end{center}
\end{table}

\begin{figure}[!htb]
	\centering
	\includegraphics[width=7.5 cm]{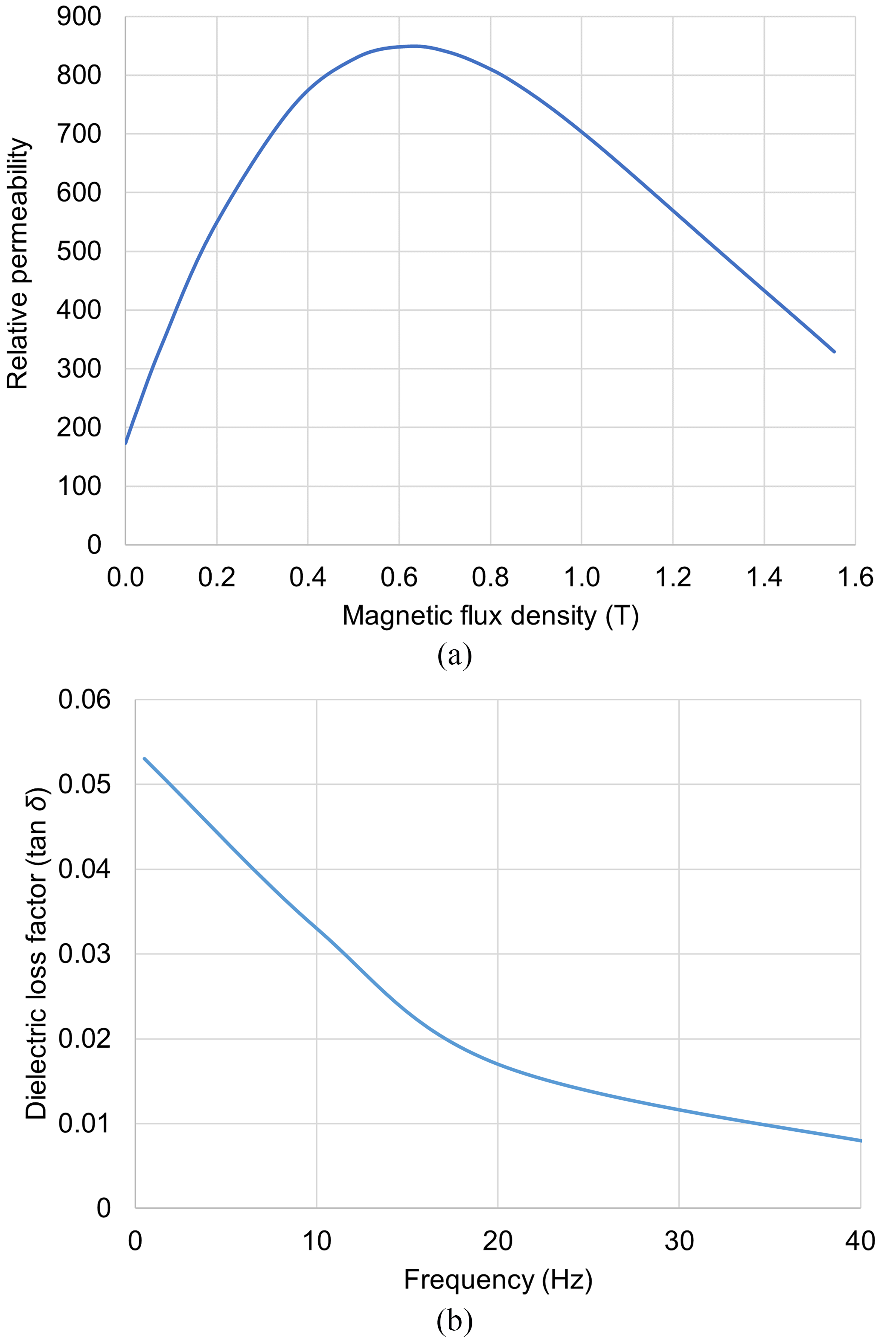}
	\caption{(a) Relative permeability for a LG steel (three cables), and (b) dielectric loss factor as a function of frequency (C2 cable).\label{fig5}}
\end{figure}

As can be seen from Table \ref{tab2}, cables with different cross section ($S_n$) and rated voltage ($V_r$), made of copper (Cu) and aluminum (Al) conductors are considered. All of them are lead sheathed, using contralay twisting for the armor wires and the power cores. Regarding the material properties, the references of Table \ref{tab2}  provide the electrical conductivity for the conductors ($\sigma_c$), sheaths ($\sigma_s$) and armor wires ($\sigma_a$) at their temperature during tests, being in most of the cases the ambient temperature ($T_{amb}$), since these are usually performed during short periods to avoid thermal effects.  On the other hand, for the three cables a typical low grade steel (LG) \cite{Hatlo2014b,Hatlo2015b} is considered for the armor wires, with the non-linear permeability ($\mu_r$) shown in Figure \ref{fig5}a. In all the cases  $\mu_r$ is supposed to be unaffected by the frequency. Regarding the relative permittivity, a complex form $\varepsilon_r($tan$\ \delta+j)$ has been considered for the C2 cable, with a frequency dependent dielectric loss factor tan $\delta$  (Figure \ref{fig5}b) \cite{Geng2018,Benato2021}. Finally, Table \ref{tab2} also provides the length of the cable $L_{cable}$ that was tested according to the references, as well as the length $L$ of the 3D geometry to be simulated in the SUSM. The difference of several orders of magnitude between the actual cable length and the length required for the simulation is remarkable.

In the following, the results presented in \cite{Arentsen2017} for the C1 cable are employed for evaluating the performance of the SUSM regarding the computation of the total losses, series impedance and induced sheath current as a function of frequency. Then, the measurements obtained in \cite{Palone2019,Benato2021} for the C2 cable are later considered for this evaluation in terms of resonant frequencies and harmonic impedances. Finally, the measurements taken by the authors of this work during the tests presented in \cite{Hatlo2014b,Hatlo2015b} are employed to evaluate the performance in terms of the MF distribution around the C3 cable as a function of frequency. In all the cases, sheaths and armor are assumed to be in SB configuration. Cables C1 and C3 were tested in laboratory setups, and C2 in its actual location. Thus, in the simulations the surrounding medium is considered as sea for cable C2 and air for the rest. Additional information regarding the experimental setups can be found in the references of Table \ref{tab2}.

\section{Simulation vs experimental results}

\subsection{Total losses}

\begin{figure}[!htb]
	\centering
	\includegraphics[width=8.5 cm]{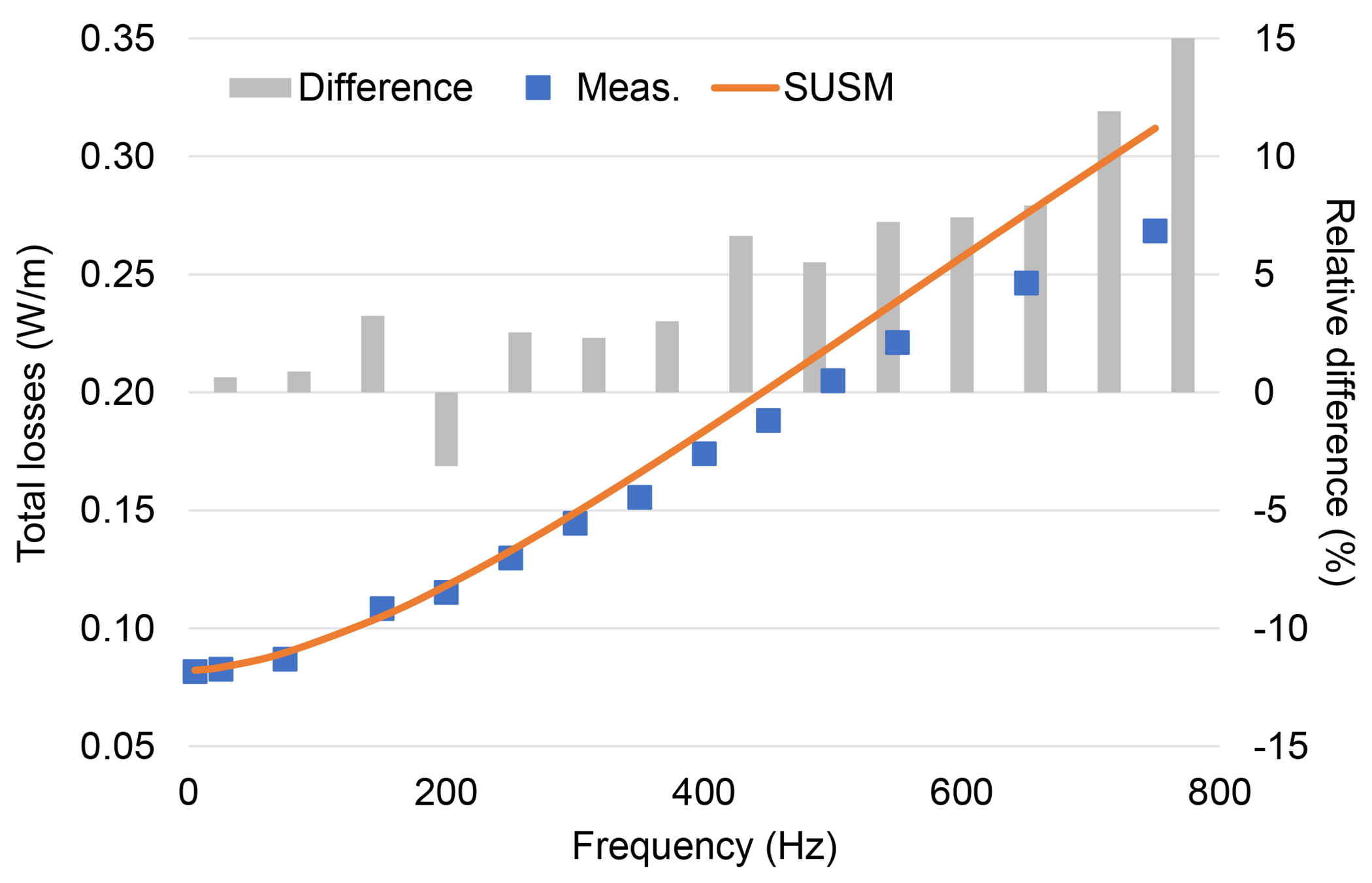}
	\caption{C1 (SB): Measured and calculated total losses for 12 A of phase current.\label{fig6}}
\end{figure}

Figure \ref{fig6} compares the measurements and computed total losses for the C1 cable (phase current of 12 A). Also, the relative differences between measurements and simulation results are represented (gray bars). As can be observed, although in most of the cases the relative differences are below 10 \%, they increase with the frequency. This is a consequence of the assumptions made in the armor dimensions and properties due to the lack of data provided by \cite{Arentsen2017}.

\subsection{Positive sequence series impedance}

\begin{figure}[!htb]
	\centering
	\includegraphics[width=8.5 cm]{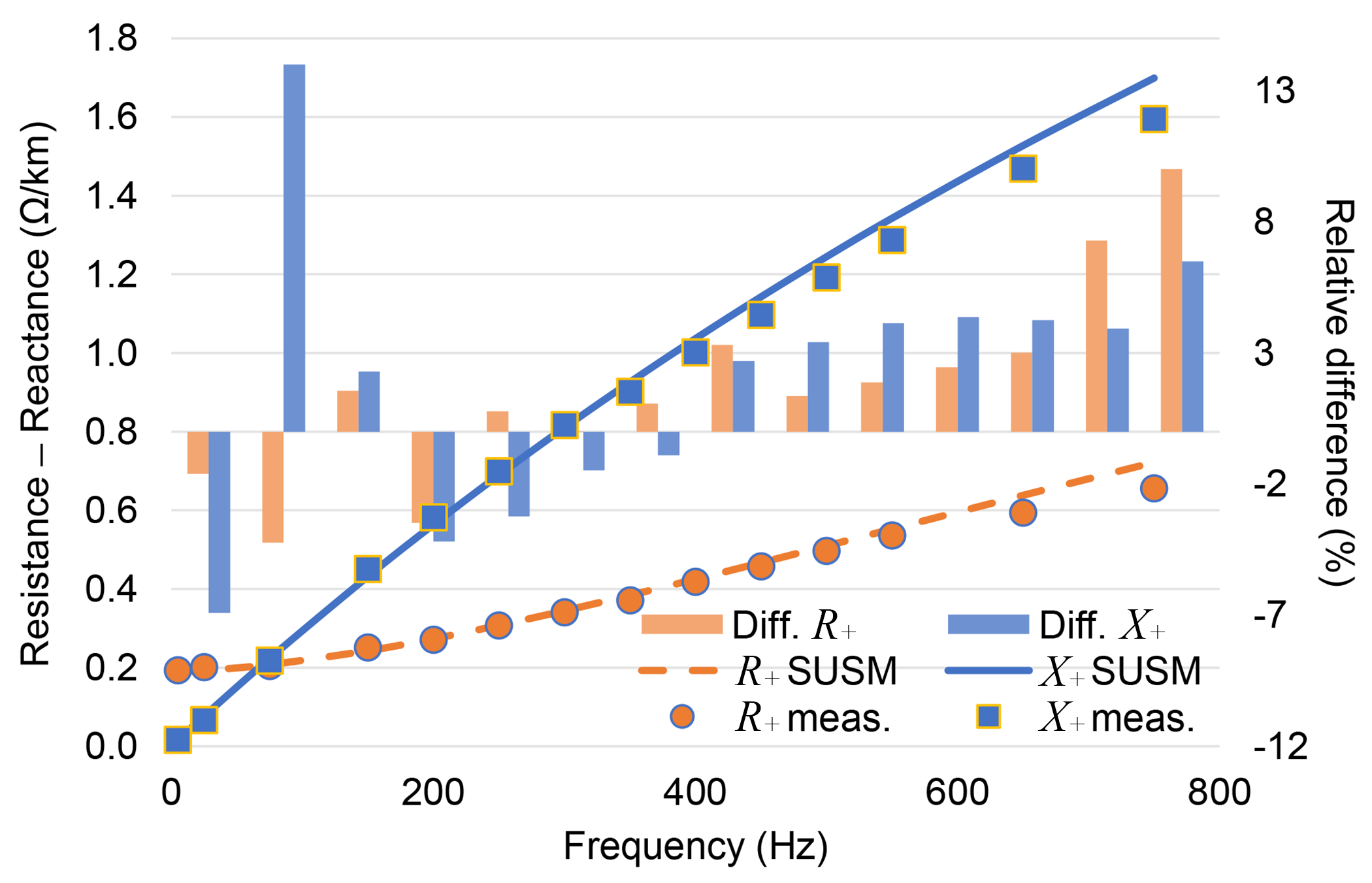}
	\caption{C1 (SB): Measured and calculated positive sequence series resistance ($R_+$) and reactance ($X_+$) for 12 A of phase current.\label{fig7}}
\end{figure}

The values of $R_+$ and $X_+$ are experimentally obtained by measuring the phase current and the total real and imaginary power involved during the test. Thus, Figure \ref{fig7} summarizes the assessment between measured and computed values for these parameters for the C1 cable, including their relative differences (color bars).

Again, increasing differences with frequency are observed in both $R_+$ and $X_+$, although they remain below 5 \% in most of the cases for the frequency range employed in the simulations. In any case, despite the tolerances in measurements and the uncertainties in the input data for the simulations, these results show a good matching between measurements and computed values derived by the SUSM.

\subsection{Sheath induced current}

\begin{figure}[!htb]
	\centering
	\includegraphics[width=8.5 cm]{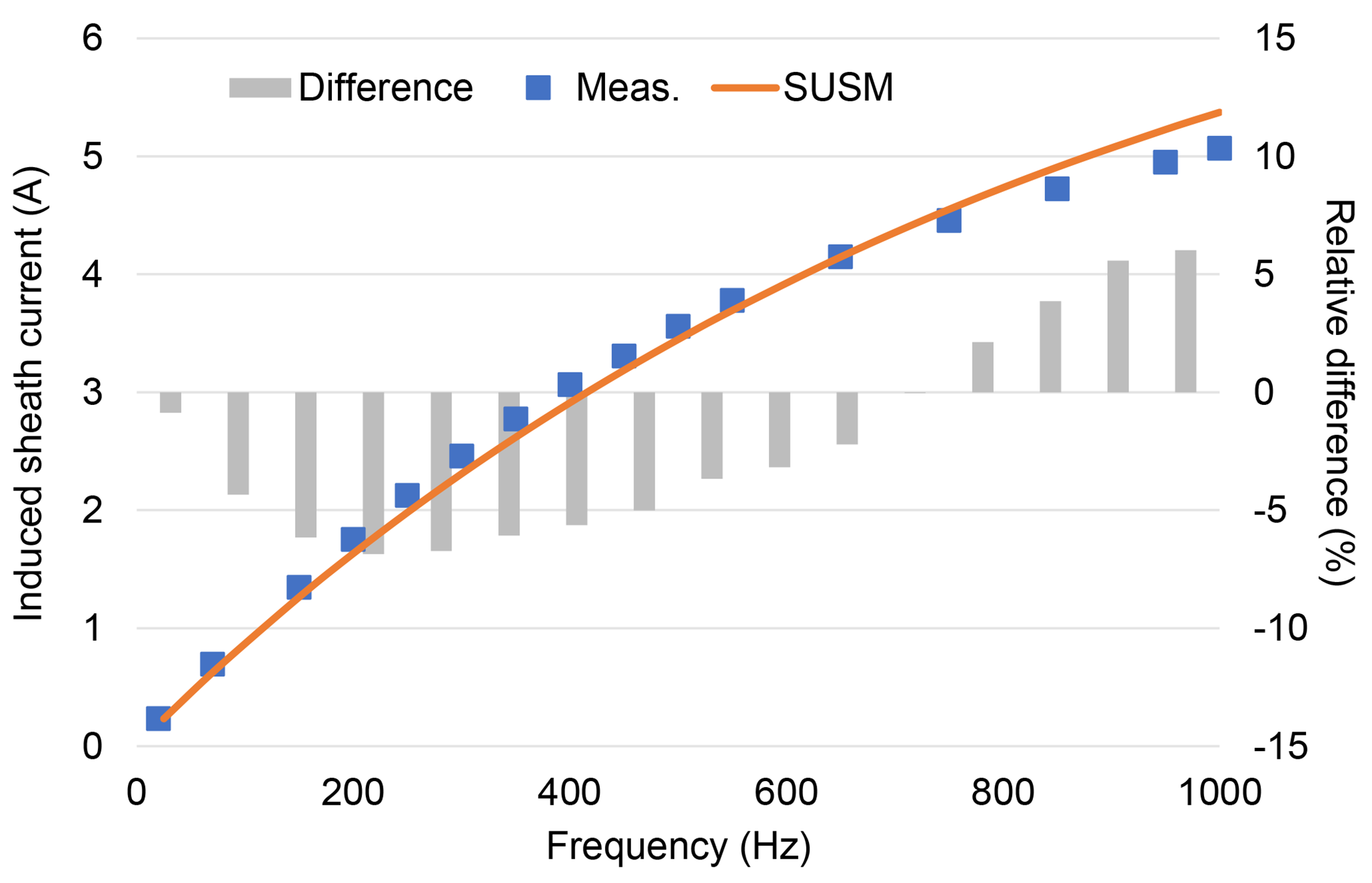}
	\caption{C1 (SB): Measured and calculated sheath current for 10 A of phase current.\label{fig8}}
\end{figure}

As can be observed in Figure \ref{fig8}, for the C1 cable there is also a very good agreement between measurements and computed values regarding the sheath induced current ($I_s$) (phase current of 10 A), since the relative differences are always well below 10 \% for all the frequencies considered in this study. Also, it can be seen that, for the highest frequencies, the computed values are greater than the measured ones. Moreover, these good results are also observed for the C3 cable, as shown in Table \ref{tab3}, where the measured and computed values for $I_s$ obtained for 50 Hz (phase current of 745 A) and 120 Hz (phase current of 304 A) are summarized, together with their relative differences. As observed, remarkable results are obtained in this cable also.

\vspace{-0.3cm}
\begin{table}[!htb]
		\renewcommand{\arraystretch}{1.1}
		\begin{center}
		\caption{C3 (SB): Measured and calculated values of the induced sheath current $I_s$.}
		\label{tab3}
			\begin{tabular}{ccc}
				& 50 Hz (745 A) & 120 Hz (304 A)\\ \hline
				Measured (A) & $187$ & $136$  \\  
				SUSM (A) & $186.3$ & $138.9$  \\ 
				Difference (\%) & $-0.37$ & $2.13$  \\ 
			\end{tabular}
	\end{center}
\end{table}

\vspace{-0.5cm}
\subsection{Harmonic impedance and resonant frequencies}

Few studies can be found in the literature providing a complete set of data regarding harmonic impedance measurements in actual TCACs. Here we reproduce the steady-state impedance measurements obtained at different frequencies (from 0 Hz to 2000 Hz) in the 99.65 km, 220 kV submarine cable link between Sicily and Malta presented in \cite{Palone2019,Benato2021} (cable C2). Positive and zero sequence impedance measurements were obtained from the sending end (labeled as 1), while the phase conductors on the other end (labeled as 2) were set up in two different conditions: open-ended (OC) or short-circuited to ground (SC). As commented earlier, in both situations sheaths and armor were in SB configuration, assuming the return current to flow through the sheaths, armor and sea during the computation of the zero sequence impedance. 

To obtain the positive and zero sequence harmonic impedances, a two-steps procedure is applied:

\begin{itemize}
	\item In a ﬁrst step, two SUSMs  are developed to obtain the per-unit length cable parameters for each frequency. The first model solves the EF in order to obtain $C$, which results in $156.25$ nF/km for the C2 cable (identical to the $156$ nF/km provided by \cite{Palone2014}). Then, the p.u. length shunt admittance (the same for both positive and zero sequences) is obtained for different frequencies through $y(f)=2\pi f C$. The second model solves the MF problem, and is supplied with either positive or zero sequence currents at different frequencies for obtaining the p.u. length positive and zero sequence impedance ($z_s(f)$, being the subscript $s=º "+"$ for the positive sequence impedance and $s=\ "0"$ for the zero sequence).
	\item In a second step, previous results are employed to obtain, from the receiving end, the positive and zero sequence harmonic impedances of the link for both OC and SC conditions. Thus, for each frequency, the exact $\pi$-model of the TCAC is expressed through the ABCD transfer matrix \cite{Xiao2012,TB556,Benato2021} as
	
	\begin{equation}\label{eq2}
		\begin{bmatrix}
			\overline{U}_{1}(f) \\
			\overline{I}_{1}(f)
		\end{bmatrix}=
		\begin{bmatrix}
			\cosh \theta_s(f) & z^c_{s}(f) \sinh \theta_s(f) \\
			\cfrac{\sinh \theta_s(f)}{z^c_{s}(f)} & \cosh \theta_s(f)
		\end{bmatrix}
		\begin{bmatrix}
			\overline{U}_{2}(f) \\
			\overline{I}_{2}(f)
		\end{bmatrix}
	\end{equation}
	
	with 
	
	\begin{equation}
		z^c_{s}(f)=\sqrt{\cfrac{z_s(f)}{y(f)}}	
	\end{equation}
	
	\begin{equation}
		\theta_s(f)=\gamma_s(f) l=\sqrt{z_s(f) \cdot y(f)} l	
	\end{equation}
	
	\noindent being $l$ the TCAC length, $\overline{U}_{1}$, $\overline{I}_{1}$, $\overline{U}_{2}$ and $\overline{I}_{2}$ the sending and receiving end voltages and currents, respectively,  $z^c_{s}$ the characteristic  harmonic impedance, and $\gamma_s$ the propagation coefficient for the frequency and sequence considered. Then, the harmonic impedances can be obtained by considering $\overline{U}_{2}(f)=0$ (SC) and $\overline{I}_{2}(f)=0$ (OC), leading to
	
	\begin{equation}\label{eq3}
		Z_{SCs}(f)=z^c_{s}(f) \tanh \theta_s(f)
	\end{equation}
	
	\begin{equation}\label{eq4}
		Z_{OCs}(f)=\cfrac{z^c_{s}(f)}{\tanh \theta_s(f)} 
	\end{equation}
	
\end{itemize}

\subsubsection{Receiving end short-circuited}

\begin{figure}[!htb]
	\centering
	\includegraphics[width=8.5 cm]{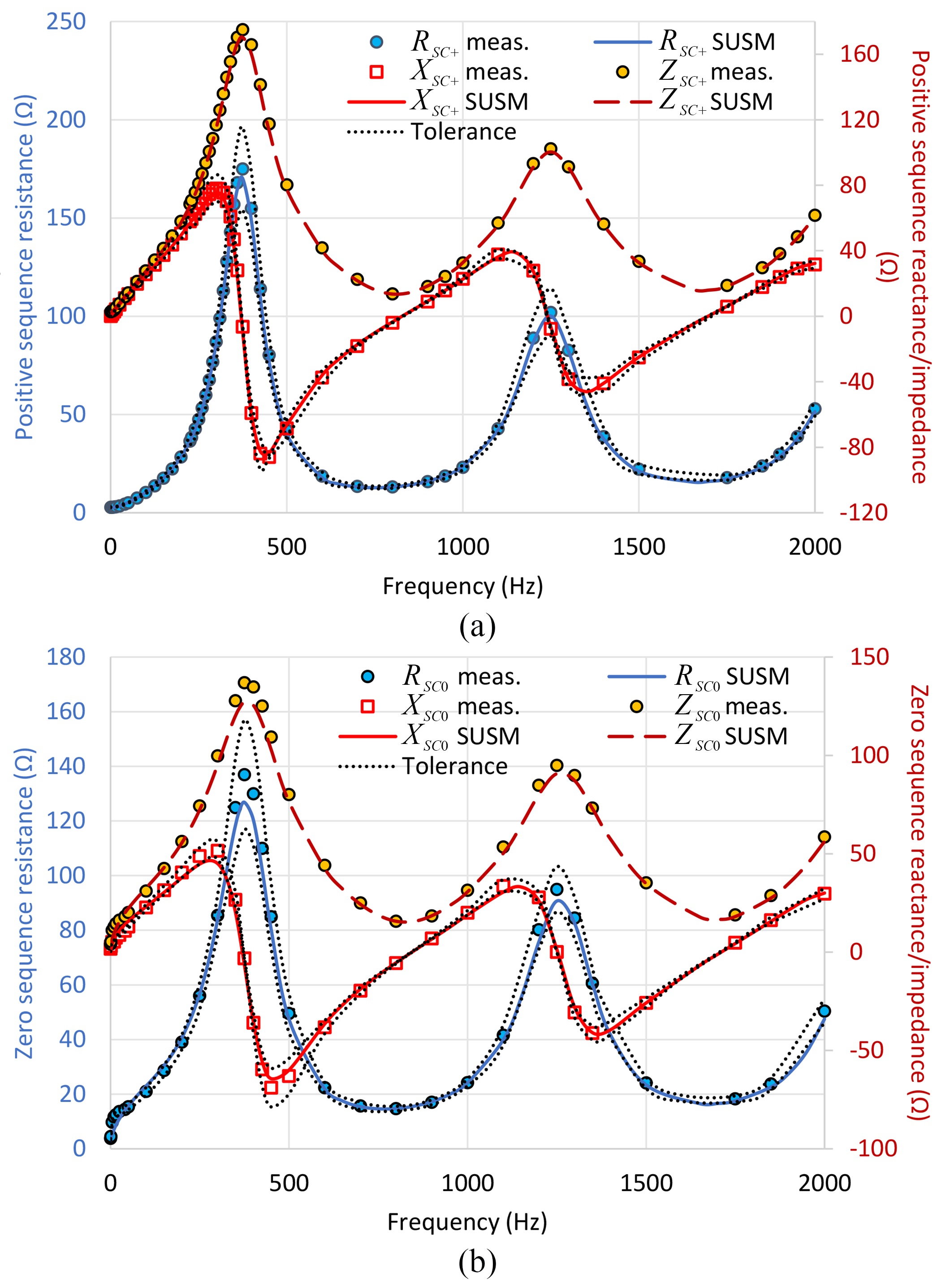}
	\caption {C2 (SB): Measured and calculated values of $R_{SC}$, $X_{SC}$ and $Z_{SC}$ for (a) the positive and (b) the zero sequence (receiving end short-circuited).\label{fig9}}
\end{figure}

For this case, Figures \ref{fig9}a and \ref{fig9}b compare the measurements and computed values of $Z_{SC+}$ and $Z_{SC0}$ together with those of their real ($R_{SC+}$, $R_{SC0}$) and imaginary ($X_{SC+}$, $X_{SC0}$) parts, including the measurement tolerances reported by \cite{Benato2021}. As can be observed, there is a very good match between both sets of data, especially for the positive sequence parameters, since differences are well below 10 \% in most of the cases, being the simulation results always between the measurement uncertainties. However, the SUSM values for $R_{SC0}$, $X_{SC0}$ and $Z_{SC0}$ give rise to errors up to 30 \% when $f<30$ Hz. This is a consequence of the simplifying assumptions  regarding the replacement of the armor wires by steel shells, as commented earlier. Nonetheless, it only occurs in a very small portion (2 \%) of the whole frequency range considered, being the simulation results always between the measurement uncertainties in the remaining cases, with maximum errors well below 10 \%.
	

\subsubsection{Receiving end open}

\begin{figure}[!htb]
	\centering
	\includegraphics[width=8.5 cm]{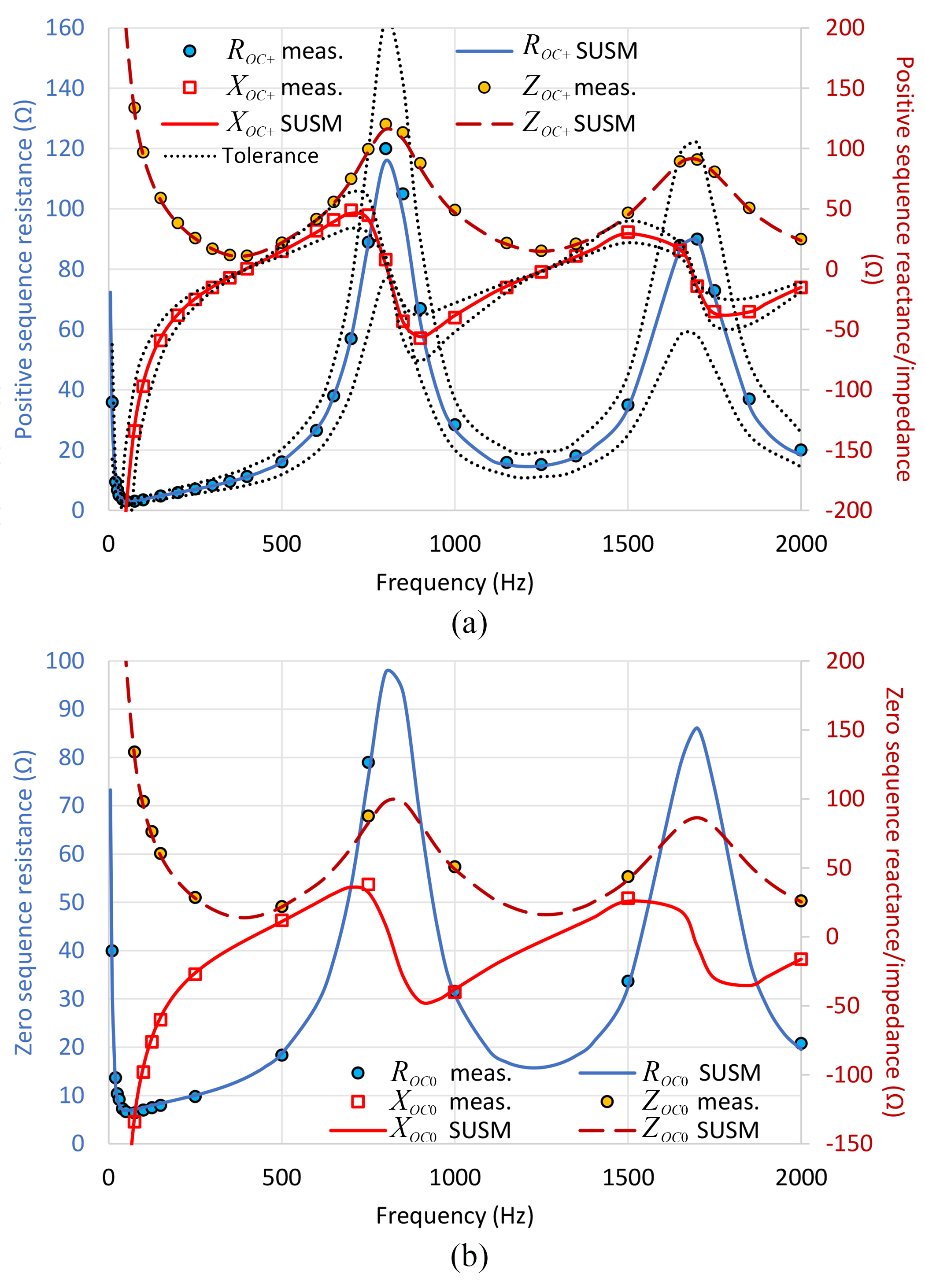}
	\caption{C2 (SB): Measured and calculated values of $R_{OC}$, $X_{OC}$ and $Z_{OC}$ for (a) the positive and (b) the zero sequence (receiving end open).\label{fig10}}
\end{figure}

In this case, results are similar to those observed in the previous section. In particular, Figures \ref{fig10}a and \ref{fig10}b represent the values for the positive ($Z_{OC+}$) and zero sequence ($Z_{OC0}$) impedances together with those of their real ($R_{OC+}$, $R_{OC0}$) and imaginary ($X_{OC+}$, $X_{OC0}$) parts (note that few measurements were taken for the zero sequence parameters during the tests, as reported in \cite{Palone2019}). A good match can be observed in the results, even at low frequencies for the zero sequence parameters, with relative differences between measurements and computed values below 10 \% in all the cases, and being these differences within the tolerance of measurement errors.

\subsubsection{Resonant frequencies}

From previous results, the resonant frequencies for both receiving end configurations can be derived. In particular, Table \ref{tab5} summarizes the measured and computed frequencies that give rise to resonances for the positive sequence impedances (peaks/valleys in $Z_{OC+}$ and $Z_{SC+}$, or zero crossings in $X_{OC+}$ and $X_{SC+}$). Table \ref{tab5} also includes its relative differences, with a maximum of about 1 \% in the worst case.

\begin{table}[!htb]
	\renewcommand{\arraystretch}{1.1}
		\begin{center}
		\caption{C2 (SB): Measured and calculated resonance frequencies.}
		\label{tab5}
			\begin{tabular}{ccc|ccc}
				\multicolumn{3}{c|}{Freq. (Hz) (SC)} & 	\multicolumn{3}{c}{Freq. (Hz) (OC)}\\ 
				Meas. & SUSM & Diff. (\%) & Meas. & SUSM & Diff. (\%) \\ \hline
				$373.5$ & $369.52$ & $-1.07$ & $392.5$ & $392$ & $-0.13$ \\  
				$830$ & $829$ & $-0.12$ & $805$ & $812$ & $0.93$ \\ 
				$1239.5$ & $1237.5$ & $-0.16$ & $1266$ & $1262.5$ & $-0.28$ \\ 
				$1705$ & $1695$ & $-0.59$ & $1675.5$ & $1672.5$ & $-0.18$ \\  
			\end{tabular}
	\end{center}
\end{table}

\vspace{-0.3cm}
From this analysis it is concluded that these results are sensitive to parameters such as $\mu_r$. In particular, better results have been obtained when considering the non-linear $\mu_r$ shown in Figure \ref{fig5}a instead of using the value of $\mu_r=300$ provided by \cite{Benato2021}.

\subsection{Magnetic field distribution}

\begin{figure}[!htb]
	\centering
	\includegraphics[width=8.5 cm]{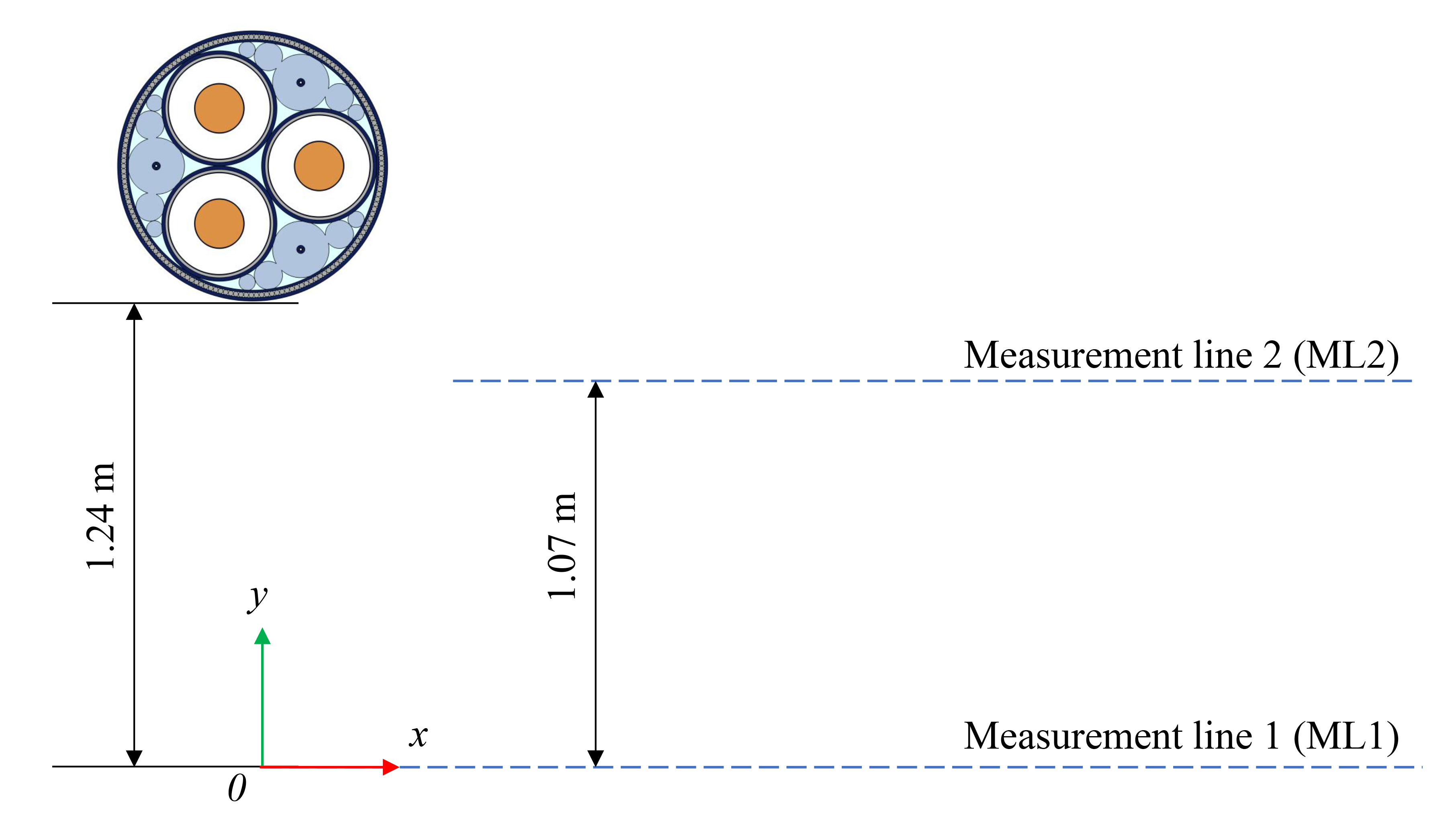}
	\caption{Measurement lines for the MF distribution.\label{fig11}}
\end{figure}

\begin{figure}[!htb]
	\centering
	\includegraphics[width=8.5 cm]{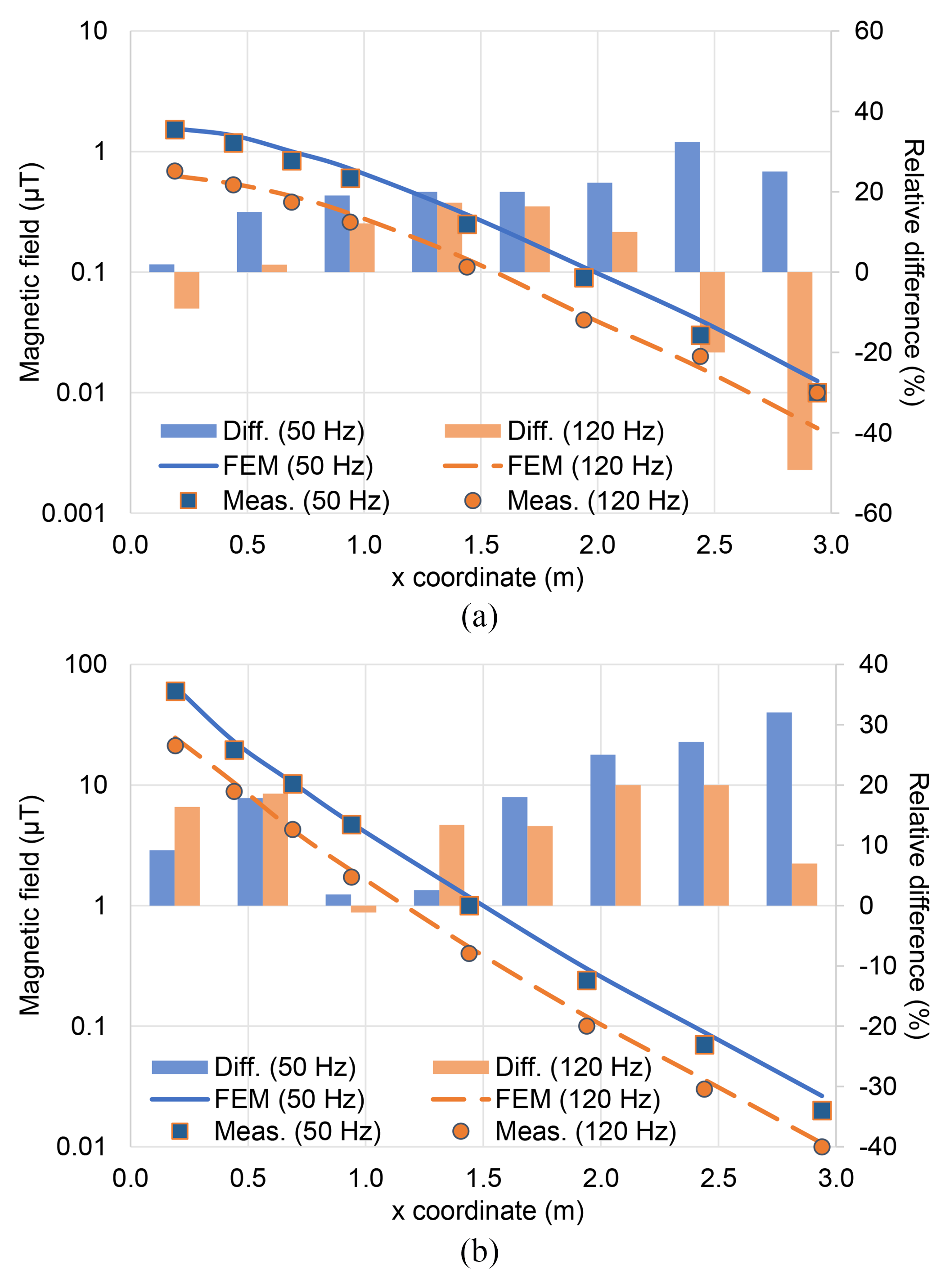}
	\caption{C3 (SB): Evolution of the measured and computed MF with the distance in (a) ML1 and (b) ML2.\label{fig12}}
\end{figure}

During the tests developed by \cite{Hatlo2014b} on cable C3, the authors of this work had the opportunity of taking MF measurements. The TCAC was suspended at 1.24 m above the ground surface, so measurements were taken at different distances from the cable axis and at different heights from the ground level (measurement lines 1 (ML1) and 2 (ML2) in Figure \ref{fig11}). A 3-axis EMDEX II MF meter was employed with a resolution of 0.01 \textmu T in the range of 0.01 to 300 \textmu T at power frequency. The measured and computed MF values obtained for 50 Hz (phase current of 745 A) and 120 Hz (phase current of 304 A) in ML1 and ML2 are represented in Figures \ref{fig12}a and \ref{fig12}b, as well as their relative differences, considering sheaths in SB (logarithmic scale employed in the MF axis for better visualization). In this case, the TCAC is assumed to be surrounded by air during the simulations.

As can be observed, good results are provided by the SUSM for both frequencies, with relative differences below 20 \% in most of the cases, although they increases with the distance. These differences are caused by a combination of two main factors. On one side, at far distances from the TCAC the resolution of the MF meter (0.01 \textmu T) is in the same order of the MF levels to be measured, leading to greater errors. On the other hand, as reported in \cite{Hatlo2014b}, a small imbalance was observed in the phase currents during the tests, giving rise to a return current through the armor of about 3 A. This disturbs the MF distribution around the TCAC, especially in areas close to the cable surface. We concluded this after some tests developed in the SUSM under different unbalanced conditions, although it was not possible to replicate the imbalance observed during the test due to a lack of data.

\section{Time response in cable energization}

\begin{figure}[!htb]
	\centering
	\includegraphics[width=8.5 cm]{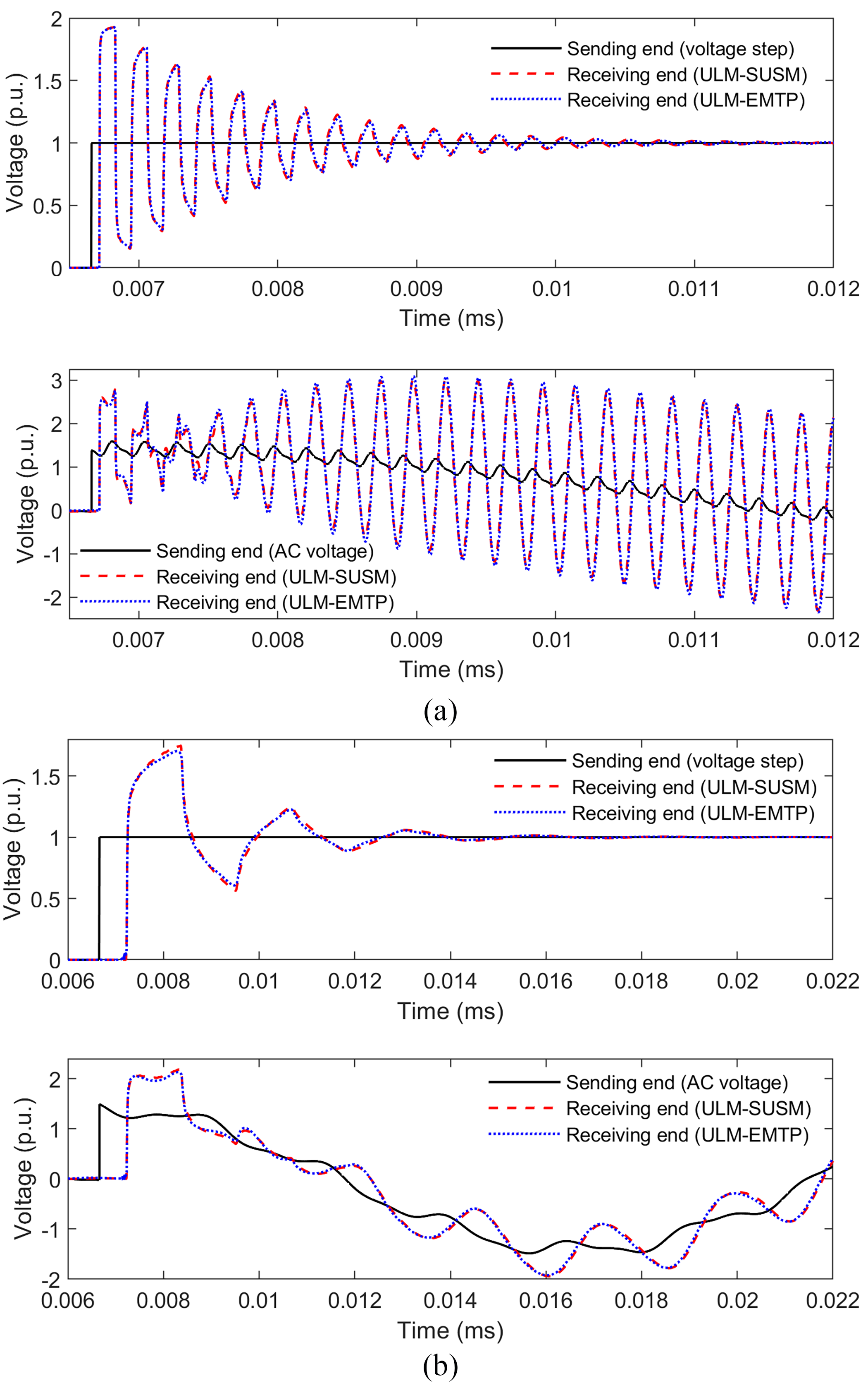}
	\caption{{C2 (SB): Time response at the receiving end when energizing the cable with a step and a sinusoidal voltage for a cable length of (a) 10 km and (b) 99.65 km}.\label{fig13}}
\end{figure}

{In order to validate the results derived from the SUSM, this section analyzes the time response in cable C2 when one of the phases is energized with both a step voltage and an AC voltage of 127 kV (rms) (the energization takes place at the peak voltage of the input source \cite{dasilva2013b,Patel2016}). This is done for two cable lengths (10 km and 99.65 km), where it is assumed that the cable is surrounded by sea water, the receiving end is open, and the sheaths and armor are in SB. The time response is computed through an universal line model (ULM) \cite{Morched99,Ramos2010}, which is derived for each cable length from the per-unit length cable parameters obtained previously through the SUSM. For testing purposes, the AC waveform includes the first and second resonant frequencies for each cable length (Table \ref{tab6}), having an amplitude of $0.1$ p.u and $0.05$ p.u., respectively. The voltage at the receiving end derived from the ULM and SUSM data (denoted as ULM-SUSM) is represented in Figures \ref{fig13}a and \ref{fig13}b together with that obtained through the ULM implemented in the software EMTP-RV \cite{emtp} (denoted as ULM-EMTP).}

\vspace{-0.3cm}
\begin{table}[!htb]
	\begin{center}
		\caption{{C2 (SB): The first four resonant frequencies for a cable length of $10$ km and $99.65$ km (receiving end open)}.}
		\label{tab6}
		\color{red}\begin{tabular}{ccccc}
			& 	\multicolumn{4}{c}{Resonant frequencies (Hz)} \\
			\multicolumn{1}{c|}{Length (km)} & $1^{st}$  & $2^{nd}$  & $3^{rd}$ & $4^{th}$  \\ \hline
			\multicolumn{1}{c|}{$10$} & $4266$ & $8611$ & $1.29\cdot 10^4$  &	$1.73\cdot 10^4$ \\ 
			\multicolumn{1}{c|}{$99.65$} &  $392$ & $812$ & $1262$ & $1672$ \\  
		\end{tabular}
	\end{center}
\end{table}

\vspace{-0.3cm}

{As can be seen, the time-response is as expected for this type of test \cite{dasilva2013b,Patel2016}, showing how a longer cable increases the damping in the waveforms. The effect of the resonant frequencies included in the input voltage is also evident in terms of harmonics amplification. Moreover, the results derived from ULM-SUSM and ULM-EMTP are very similar, with deviations up to 7 \% in the waveform amplitude. So from these preliminary results it can be concluded that the SUSM is also able to provide enough and accurate information for the analysis of electromagnetic transients in TCACs, although this deserves an in-depth analysis in a further study due to the relevance of this topic.}


\color{black}
\section{Conclusions}

This paper confronts, for the first time in three-core armored cables, 3D-FEM simulation results with experimental measurements in the harmonic frequencies. To this aim, a new SUSM is proposed that reduces simulation time in about 30 \% when performing frequency sweeps.

Different frequency-domain analyses are performed through the SUSM for evaluating relevant aspects in three real TCACs, such as total losses, series resistance, inductive reactance and induced sheath current, obtaining relative differences well below 10 \% between measurements and computed values. Also, the SUSM has proved to provide very accurate results when computing the sequence impedances in the harmonic frequencies, with relative differences typically below 5 \% in the positive sequence impedance in both OC and SC configurations. However, and just when computing the zero sequence impedance for the SC case, the SUSM loses accuracy (differences below 30 \%) in a very limited range of frequencies ($f<$ 30 Hz). Despite this fact, resonant frequencies are accurately estimated, with differences below 1 \%. In all cases the results are, at least, as accurate as other alternative methods based on 3D geometries \cite{Benato2021}. Also, remarkable results are obtained when estimating the MF distribution around TCACs, with differences below 20 \% when compared to measurements. {Eventually, a simple application to switching transient analysis has been presented, leading to similar results to those obtained with the electromagnetic transients program EMTP-RV.}

As a major conclusion, even considering the uncertainties in measurements and input data, and the lack of more published results from harmonic measurements, remarkable results are obtained, presenting the SUSM as a valuable tool for assessing frequency-domain analyses of TCACs and a perfect complement to costly experimental setups.

%

\section*{Acknowledgment}
The authors would like to acknowledge and thank Jarle Bremnes and Marius Hatlo  for bringing the opportunity to take MF measurements during their tests in the Halden factory.

\ifCLASSOPTIONcaptionsoff
  \newpage
\fi


\bibliography{IEEEabrv,mybibfile}

%








\end{document}